\documentclass{article}

\usepackage{graphicx}

\def\abstract#1{\vskip 7mm 
        \begin{center}{\large Abstract}\par \smallskip
                \begin{minipage}[c]{12cm}
                        \small #1
                \end{minipage}
        \end{center}
}
\def\title#1{\begin{center}{\Large\bf #1}\end{center}}
\def\author#1{\vskip 5mm \begin{center}{#1}\end{center}}
\def\address#1{\begin{center}{\it #1}\end{center}}
\def\az{\!\stackrel{\hbox{\tiny (0)}}{a}\!\!}

\def\Hz
{\!\stackrel{\hbox{\tiny (0)}}{\cal H}\!{}}
\def\dH{{\delta {\cal H}}}
\def\phiz
{\!\stackrel{\hbox{\tiny (0)}}{\phi}\!{}}
\def\tildephiz
{\!\stackrel{\hbox{
$\sim\hspace{-11pt}{\lower3pt\hbox{\tiny(0)}}$}}\phi\!{}}

\def\Ez
{\!\stackrel{\hbox{\tiny (0)}}{\cal E}\!{}}
\makeatletter
\@ifundefined{lesssim}{\def\lesssim{\mathrel{\mathpalette\vereq<}}}{}
\@ifundefined{gtrsim}{}{}
\def\vereq#1#2{\lower3pt\vbox{\baselineskip1.5pt \lineskip1.5pt
\ialign{$\m@th#1\hfill##\hfil$\crcr#2\crcr\sim\crcr}}}
\makeatother

\begin{document}

\title{%
  Leading order corrections to the cosmological evolution
  of tensor perturbations in braneworld
}
\author{%
  Tsutomu Kobayashi\footnote{E-mail:tsutomu@tap.scphys.kyoto-u.ac.jp} and
  Takahiro Tanaka\footnote{E-mail:tama@scphys.kyoto-u.ac.jp}
}
\address{%
  Theoretical Astrophysics Group,
  Department of Physics, Kyoto University,
  Kyoto 606-8502, Japan
}

\abstract{
Tensor perturbations in an expanding 
braneworld of the Randall Sundrum type are investigated. 
We consider a model composed of a slow-roll inflation phase 
and the succeeding radiation phase. The effect of the 
presence of an extra dimension through the transition 
to the radiation phase is 
studied, giving an analytic formula for leading 
order corrections. 
}

\section{Introduction}

In recent years there has been much interest in braneworld
scenarios~\cite{Maartens:2003tw}.
Among various models, the second Randall-Sundrum model (RS II)~\cite{Randall:1999vf}
with one brane in an anti de Sitter (AdS) bulk
is particularly interesting
in the point that, despite an infinite extra dimension,
four dimensional general relativity (4D GR) can be recovered
at low energies/long distances on the brane.
Then, what is leading order corrections to
the conventional gravitational theory?
It was shown that
the Newtonian potential in the RS II braneworld,
including the correction due to the bulk gravitational effects
with a precise numerical factor, is given by~\cite{Garriga:1999yh}
\begin{eqnarray*}
V(r)\simeq -\frac{Gmm'}{r}\left(1+\frac{2\ell^2}{3r^2}\right).
\end{eqnarray*}
Here $\ell$ is the curvature length of the AdS space
and is experimentally constrained to be $\ell \lesssim 0.1$ mm.
It is natural to ask next what are leading order corrections
to the evolution of cosmological perturbations.

Great efforts have been paid for the problem of
calculating cosmological perturbations in braneworld scenarios.
In order to correctly evaluate perturbations on the brane,
we need to solve bulk perturbations, which reduces to a problem
of solving partial differential equations with appropriate 
boundary conditions.
There is another difficulty concerned with a physical, fundamental aspect
of the problem; we do not know how to specify appropriate initial conditions
for perturbations with bulk degrees of freedom.
Hence it is not as easy as in the standard four dimensional cosmology.

In this short article,
we investigate cosmological tensor perturbations in
the RS II braneworld, and evaluate leading order corrections
to the 4D GR result analytically.
For this purpose, 
we make use of the reduction scheme to a four dimensional 
effective equation which
iteratively takes into account the effects of the bulk gravitational
fields~\cite{Tanaka:2004ig}. 
As a first step, we concentrate on perturbations 
which is initially at super-horizon scales. 
A detailed discussion is done in a much longer paper~\cite{Kobayashi:2004wy}.

\section{Tensor perturbations on a Friedmann brane}

\paragraph{Background model:}
The background spacetime that we consider is composed of 
a five dimensional AdS bulk, whose metric is
given in Poincar\'{e} coordinates as
$
ds^2=(\ell/z)^2 \left(
dz^2-dt^2+\delta_{ij}dx^idx^j \right)
$,
with a Friedmann brane
at $z=z(t)$.
The induced metric on $z=z(t)$ is
$
ds^2=a^2(t)\left[-(1-\dot z^2)dt^2+\delta_{ij}dx^idx^j\right],
$
where $a(t)=\ell / z(t)$ and the overdot denotes $\partial_t$.
From this we see that the conformal time on the brane
is given by $d\eta = \sqrt{1-\dot z^2} dt$.
The brane motion is related to the energy density of matter 
localized on the brane by the modified Friedmann
equation as
\begin{eqnarray}
{\cal H}^2=\frac{8\pi G}{3} a^2\left( \rho + \frac{\rho^2}{2\sigma} \right),
~~{\cal H}=-\dot z/(z \sqrt{1-\dot z^2}),
\label{FriedmannEq}
\end{eqnarray}
where $\sigma =3/4\pi G \ell^2$ is a tension of the brane.
The quadratic term in $\rho$
in the Friedmann equation
modifies the standard cosmological expansion law.
As a background Friedmann brane model, we consider 
slow-roll inflation at low energies,
characterized by a small slow-roll parameter,
$\epsilon:=1-\partial_{\eta}{\cal H}/{\cal H}^2\ll 1$,
followed by a radiation dominant phase, where $a\propto \eta +{\cal O}(\ell^2)$.


\paragraph{Tensor perturbations on the brane:}
We use the method to reduce the five dimensional equation 
for tensor perturbations in a Friedmann 
braneworld
at low energies to a four dimensional effective equation of motion, 
derived in Ref.~\cite{Tanaka:2004ig}.
Tensor perturbations on a Friedmann
brane
are given by
$
ds^2=(\ell/z)^2 \left[
dz^2-dt^2+(\delta_{ij}+h_{ij})dx^idx^j \right].
$
We expand the perturbations by
using $Y^k_{ij}(x)$, a transverse traceless tensor harmonics with 
comoving wave number $k$, 
as $h_{ij}=\sum_{k}Y^k_{ij}(x)\Phi_k(t,z)$. 
Then the equation of motion for the tensor perturbations in the bulk 
is given by
\begin{eqnarray}
\left( - \partial_z^2+\frac{3}{z} \partial_z+\partial_t^2+k^2 \right)
\Phi_k=0,
\end{eqnarray}
Hereafter we will discuss each Fourier mode separately, 
and we will abbreviate the subscript $k$.
The general solution of this equation is 
$
\Phi
=\int d\omega \tilde \Psi(\omega) e^{-i\omega t}
2(pz)^2K_2(pz)
$
where $p^2=-\omega^2+k^2$. We have chosen
the branch cut of the modified Bessel function $K_2$ so that
there is no incoming wave from past null infinity in the bulk.
The coefficients $\tilde \Psi(\omega)$ are to be determined
by the boundary condition on the brane
$n^{\mu}\partial_{\mu}\Phi|_{z=z(t)}=0$, where $n^{\mu}$
is an unit normal to the brane, or, equivalently,
by the effective Einstein equations on the brane~\cite{Shiromizu:1999wj},
\begin{eqnarray}
^{(4)}G_{\mu\nu}=8\pi G\:T_{\mu\nu}+(8\pi G_5)^2 \pi_{\mu\nu}
-E_{\mu\nu}.
\end{eqnarray}
A projected Weyl tensor $E_{\mu\nu}:=C_{\alpha\mu\beta\nu}n^{\alpha}n^{\beta}$
represents the effects of
the bulk gravitational fields, giving rise to corrections to
four dimensional Einstein gravity in a fairly nontrivial way.
In the present case, this can be written explicitly.
After some manipulation
the effective four dimensional equation reduces to
\begin{eqnarray}
\left( \partial_{\eta}^2+2\Hz \partial_{\eta}+k^2\right) \phi
\simeq \frac{\ell^2}{\az^2}\left(
 {\cal S}_0[\phi]+{\cal S}_1[\phi] + {\cal S}_2[\phi] 
\right),
\label{basic}
\end{eqnarray}
where
\begin{eqnarray}
{\cal S}_0[\phi]=-\frac{2\az^2}{\ell^2}\dH \phi', 
~
{\cal S}_1[\phi]=
(3\Hz ^3-2\Hz \Hz')\phi' +k^2\Hz ^2,
~
{\cal S}_2[\phi]=-\frac{1}{2}\int d\omega
p^4\tilde \phi e^{-i\omega \eta}
\left[ \ln \left(\frac{p\ell}{2a}\right) +\gamma \right],
\end{eqnarray}
and the prime denotes $\partial_{\eta}$.
This is our basic equation.
The cosmic expansion rate at zeroth order is described by $\Hz$ and $\az~$,
and can be obtained by dropping
the quadratic term in $\rho$ in Eq.~(\ref{FriedmannEq}).
The first term ${\cal S}_0$ arises due to the non-standard cosmic expansion 
included in ${\cal H}$, while the other two terms are 
the corrections from the bulk effects $E_{\mu\nu}$.
We can see that all terms 
are suppressed by $\ell^2$ (or $\ell^2\ln\ell$).
${\cal S}_2$ is essentially nonlocal because of 
the presence of the log term.

\section{Leading order corrections}

The perturbation equation~(\ref{basic})
can be solved iteratively by taking $\ell^2$ as a small parameter.
We write
$
\phi=
\phiz e^{F(\eta)}=
\phiz\exp\left[\int^{\eta}f(\eta') d\eta' \right],
$
and a zeroth order solution $\phiz$
by definition satisfies 
the equation obtained by dropping the source terms on the right hand side. 
Then $f$ obeys
$
\partial_{\eta}f+2(\partial_\eta\phiz/ \phiz+\Hz)f
= \ell^2 {\cal S}[\phiz]/\az^2 \phiz,
$
which can be integrated immediately to give
$
f(\eta) = (\ell/\az \hspace{1.5mm}\phiz)^2
\int^{\eta}
d\eta' \phiz{\cal S}[\phiz].
$
Here ${\cal S}={\cal S}_0+{\cal S}_1+{\cal S}_2$.
Integrating this expression, we obtain the first order correction $F$.
We hereafter denote the corrections $f$ and $F$ 
coming from ${\cal S}_i$ as $f_i$ and $F_i$, respectively.

During slow-roll inflation and the earlier stage of the radiation dominated phase,
the perturbation modes are well outside the horizon and
thus we can use the long wavelength approximation,
which simplifies the calculation.
In the long wavelength approximation,
the zeroth order solution can be explicitly written as
$
\phiz\simeq
A_k\left[1-k^2\int^{\eta}d\eta'I(\eta')\right],
$
where $A_k$ is an amplitude of the perturbation 
which can depend on $k$, and
$
I(\eta):={\az^{-2}(\eta)}\int^{\eta}_{-\infty} \az^2(\eta')d\eta'.
$
We simply keep the terms up to ${\cal O}(k^2)$.

\paragraph{Slow-roll inflation:}
Let us consider slow-roll inflation on the brane.
In this case ${\cal S}_0$ vanishes, since 
$\dH=0$ by construction. 
Besides,
$\hat p^4 \phiz$ is shown to vanish at first order in the slow-roll parameters
during inflation.
Consequently, the only relevant term
in the present situation is ${\cal S}_1\simeq -A_kk^2\Hz ^2\epsilon$.
Integrating twice, we obtain
\begin{eqnarray}
F(\eta) \approx \frac{k^2\ell^2}{2 \az^2}\epsilon.
\end{eqnarray}
Here we chose the integration constant so that $F$
vanishes in $a\to\infty$ limit.
We see that the correction arising during inflation
is very tiny, suppressed by the slow-roll parameter 
$\epsilon$ in addition to the factor $k^2\ell^2/\az^2$.
For pure de Sitter inflation there is no correction, as is expected.

\begin{figure}[bt]
  \begin{center}
    \includegraphics[keepaspectratio=true,height=45mm]{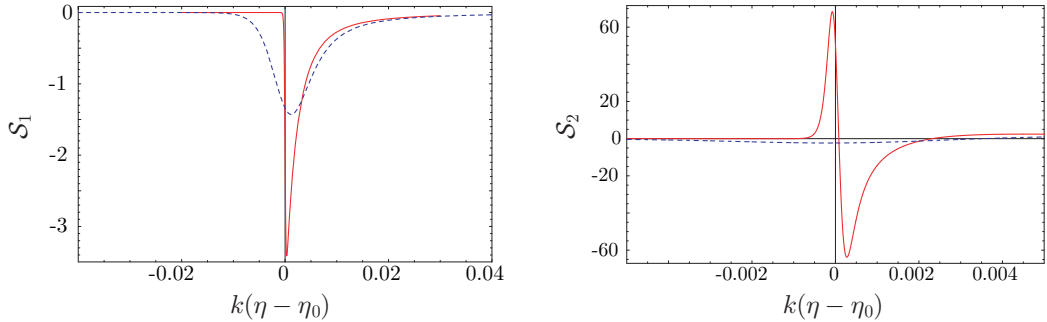}
  \end{center}
  \caption{Behavior of the source terms of corrections
  ${\cal S}_1$ and ${\cal S}_2$ around the transition time
  with the vertical axis in an arbitrary unit.
  Solid line shows the case of a sharp transition
  with $s=0.02\eta_0$, 
  while dashed line represents the case of a smooth transition
  with $s=0.5\eta_0$.
  The wavelength of the mode is chosen to be $k\eta_0=0.01$.}
  \label{fig:fig.eps}
\end{figure}

\paragraph{Transition from inflation to the radiation stage:}
Now we investigate the effects of the transition from
the inflation stage to the radiation stage.
First, just for the illustrative purpose,
we plot the behavior of ${\cal S}_1$ and ${\cal S}_2$
for the modes well outside the horizon
in the neighborhood of the transition time $\eta_0$
in Fig.~\ref{fig:fig.eps}. 
These plots are for $\epsilon$ given by
$
\epsilon(\eta)= \tanh[(\eta-\eta_0)/s]+1,
$
where the parameter $s$ controls the smoothness of the transition.
From the figures it can be seen that
the corrections from $E_{\mu\nu}$, 
${\cal S}_1$ and ${\cal S}_2$,
become significant only around the transition time.

Let us consider the limiting case where
$\epsilon$ is given by $\epsilon(\eta)=2\theta(\eta-\eta_0)$.
We neglect the tiny effect of the non vanishing 
slow-roll parameter during inflation.
In this case, by an elaborate calculation
we can analytically obtain the corrections
from the source term ${\cal S}_1$ and ${\cal S}_2$.
They become time independent long time after the transition,
and are given by
\begin{eqnarray}
\lim_{\eta\to\infty }F_1(\eta) 
=
 -\frac{k^2\ell^2}{a^2_0},~
\lim_{\eta\to \infty}F_2(\eta)=
\frac{2k^2\ell^2}{a_0^2}\left\{
\frac{37}{75}-\frac{2}{15}\left[\gamma+\ln\left(
\frac{k}{a_0H_0}
\right)\right]
+\frac{1}{5}\ln\left(\frac{\ell H_0}{2}\right)
\right\}.
\label{F12fin}
\end{eqnarray}

\paragraph{Corrections due to the unconventional cosmic expansion:}
It might be possible that
further corrections arise after the mode re-enters the horizon.
However, we can show that such corrections are highly suppressed.
A basic observation supporting this conclusion is that 
the contributions from $E_{\mu\nu}$ term in the right hand side of Eq.~(\ref{basic})
become significantly large only at the transition time.
It decreases in powers of $a(\eta)$ after the transition,
and will be negligible when the long wavelength approximation brakes down.
Hence,
the long wavelength approximation
suffices for our purpose to obtain the corrections from $E_{\mu\nu}$.

Now we compute the correction due to the unconventional cosmic expansion.
In the radiation stage,
$
u(\eta):=
        a(\eta)\phi(\eta), 
$
is a convenient variable. 
In terms of this new variable, we can rewrite the equation of motion~(\ref{basic}) as
\begin{eqnarray}
u''+k^2u
=
\frac{\ell^2}{a} \left(\bar{\cal S}_0+ {\cal S}_1+{\cal S}_2\right),
\label{ham_osc}
\end{eqnarray}
with
$
 \bar{\cal S}_0=(a''/\ell^2)u. 
$
If we neglect the deviation of the expansion law from the standard one,
we have $a\propto \eta$ in the radiation stage and so $a''=0$.
This means that, just as ${\cal S}_0$, the first term on the right hand side
gives a contribution of ${\cal O}(\ell^2)$ 
from the modification of the expansion law.
Thus, all the terms collected on the right hand side are of 
$O(\ell^2)$, 
and at zeroth order $u$ behaves like a harmonic oscillator.
We define ``energy'' of the harmonic oscillator as
${\cal E}:=|u'|^2/2+k^2|u|^2/2$.
At the zeroth order this energy is conserved, but
taking into account the correction of ${\cal O}(\ell^2)$
it varies with time due to the ``external force'' $\bar{\cal S}_0$.
The variation of the energy,
$\Delta {\cal E}_0=\int^{\infty}{{\cal E}_0}'d\eta=
\int^{\infty}[\ell^2(u')^*\bar{\cal S}_0/2a+~\mbox{c.c.}]d\eta$,
is related to that of the amplitude of the oscillator,
i.e., the amplitude of the tensor perturbations, via
$\lim_{\eta\to\infty}\{ \Re[F_0]+ \delta a/\az~ \}=\Delta {\cal E}_0/(2\Ez)$.
Using this formula it is easy to obtain the correction arising due to the modification
of the expansion law:
\begin{equation}
\lim_{\eta\to\infty}\Re[F_0(\eta)]= 
\frac{1}{10}\frac{k^2\ell^2}{a_0^2}.
\label{F0fin}
\end{equation}

\section{Summary}

We have investigated leading order corrections
to tensor perturbations in the RS II braneworld cosmology
by using the perturbative expansion scheme of Ref.~\cite{Tanaka:2004ig}.
We have studied a model composed of slow-roll inflation on the brane,
followed by a radiation dominant era. 
In our expansion scheme the asymptotic boundary 
conditions in the bulk are imposed by choosing 
outgoing solutions of bulk perturbations, whose general 
expression is known in the Poincar\'{e} coordinate system. 
Hence, the issue of bulk boundary conditions is handled without 
introducing an artificial regulator brane. 
This is one of the notable advantages of the present scheme.

Combining the results obtained in Eqs.~(\ref{F12fin}) and (\ref{F0fin}),  
we find that the amplitude of a fluctuation with comoving 
wave number $k$ is 
modified by a factor $e^{\Re[F]}$ due to the effect of 
an extra dimension
with 
\begin{eqnarray}
\lim_{\eta\to \infty}\Re[F(\eta)]=
\frac{k^2\ell^2}{a_0^2}\left\{
\frac{13}{150}-\frac{4}{15}\left[\gamma+\ln\left(
\frac{k}{a_0H_0}
\right)\right]
+\frac{2}{5}\ln\left(\frac{\ell H_0}{2}\right)
\right\}, 
\end{eqnarray}
where $a_0$ and $H_0$ are the scale factor and the Hubble 
parameter at the transition time, and $\gamma$ is the Euler's constant. 
Leading corrections are proportional to $k^2\ell^2/a_0^2$ or 
$k^2\ell^2/a_0^2\log \ell$, as is expected from the dimensional 
analysis.  
However, our calculation here determined 
the precise numerical factors analytically.



\end{document}